\documentclass[prd,preprint,nofootinbib]{revtex4}

\usepackage{graphicx}
\usepackage{amsfonts}
\usepackage{latexsym}
\usepackage{amsmath}
\usepackage{amssymb}
\usepackage{epsfig}

\begin{document}

\title{Dirac dark matter, dark radiation, and the type-II seesaw mechanism in alternative $U(1)_X$ standard model}

\author{Nobuchika Okada}
 \email{okadan@ua.edu}
 \affiliation{
Department of Physics and Astronomy, 
University of Alabama, Tuscaloosa, Alabama 35487, USA}

\author{Osamu Seto}
 \email{seto@particle.sci.hokudai.ac.jp}
\affiliation{Institute for the Advancement of Higher Education, Hokkaido University, Sapporo 060-0817, Japan}
\affiliation{Department of Physics, Hokkaido University, Sapporo 060-0810, Japan}

%

\begin{abstract}
We propose an extra $U(1)_X$ model with an alternative charge assignment for right-handed right-handed neutrinos. 
The type-II seesaw mechanism by a triplet Higgs field is promising for neutrino mass generation because of the alternative charge assignment. 
The small vacuum expectation value (VEV) of an additional Higgs doublet naturally leads to a very small VEV of the triplet Higgs field, and as a result, the smallness of neutrino mass can be understood. 
With the minimal Higgs field for $U(1)_X$ with the charge $1$, right-handed neutrinos are candidates for Dirac dark matter (DM) and dark radiation (DR). 
We have derived and imposed the LHC bound, the DR constraint and the bound from DM direct searches in the wide range of parameter space.
Among various $U(1)_X$ choices, the DM direct search bound is found to be weakest for $U(1)_R$ where the constraints from thermal DM and non-negligible DR can be compatible.
Such a number of the effective neutrino species would be interesting from the viewpoint of the so-called Hubble tension.
\end{abstract}


\preprint{EPHOU-22-006} 

\vspace*{3cm}
\maketitle


\section{Introduction}

The standard model (SM) of particle physics has been constructed on the basis of the gauge principle and
 the spontaneous symmetry breaking by the Higgs mechanism at the vacuum.
The introduction of an extra $U(1)$ gauge interaction is one of the promising and
 well-defined extensions of the SM.
The $B-L$ (baryon number minus lepton number) appears to be an accidental
 global symmetry in the SM, indicating that this might be a gauge symmetry
 in a UV completion of the theory~\cite{Pati:1973uk,Davidson:1978pm,Mohapatra:1980qe,Mohapatra:1980}.
At the same time, the cancellation of an anomaly for a chiral gauge theory is critical.
When the gauge symmetry is extended from the SM gauge group $SU(3)_C\times SU(2)_L\times U(1)_Y$ to  $SU(3)_C\times SU(2)_L\times U(1)_Y \times U(1)_{B-L}$, 
 the number of right-handed neutrinos and those charge are limited to two choices by the anomaly cancellation conditions.
One is three right-handed neutrinos with each $U(1)_{B-L}$ charge $-1$,
 which has been usually considered - and might be called - the standard assignment.
Under this assignment, it is straightforward to explain observed neutrino masses by the type-I seesaw mechanism~\cite{Minkowski:1977sc,Yanagida:1979as,GellMann:1980vs,Mohapatra:1979ia} with right-handed neutrinos,
 by introducing the scalar field with the $B-L$ charge $2$ that generates the Majorana masses of right-handed neutrinos.
In the other charge assignment, two out of three right-handed neutrinos have the $U(1)_{B-L}$ charge of $-4$
 and the one has the charge of $+5$, which is sometimes called the alternative assignment~\cite{Montero:2007cd}.

A similar extra $U(1)$ extended model can be constructed based on the $U(1)_R$ gauge symmetry
 where only right-handed fermions are charged while left-handed ones are not charged~\cite{Jung:2009jz}.
The $U(1)_R$ charged particles are same as in the $U(1)_{B-L}$ model up to chirality, 
 neutrino masses can be generated by the type-I seesaw mechanism at tree level~\cite{Nomura:2017tih}
 or loop level~\cite{Nomura:2017ezy}. 
The implication due to the chirality dependence~\cite{Jung:2009jz,Jana:2019mez,Das:2021esm}, the axial-vector coupling~\cite{Seto:2020jal}, $U(1)_R$ charged Higgs scalars~\cite{Ko:2013zsa,Ko:2012hd}, and $U(1)_R$ interacting dark matter (DM)~\cite{Chao:2017rwv,Seto:2020udg,Okada:2020evk,Nagao:2020azf}, can be found in literature.
The anomaly-free most general gauged $U(1)$ extension of the SM is
 defined as a linear combination of the SM $U(1)_Y$ and the $U(1)_{B-L}$ gauge groups. 
With this convenient parametrization of $U(1)_X$~\cite{Appelquist:2002mw,Oda:2015gna,Das:2016zue},
 we can study a wide class of extra $U(1)$ models including the representative model $U(1)_{B-L}$ and $U(1)_R$. 
In this paper, we consider $U(1)_X$ models with the alternative $U(1)_{B-L}$ assignment.

Under the alternative $B-L$ charge assignment,
 it is nontrivial for the type-I seesaw mechanism to generate neutrino masses,
 because right-handed neutrinos cannot form Yukawa coupling with lepton doublets and
 the SM Higgs doublet due to the charge mismatch~\cite{Ma:2014qra,Sanchez-Vega:2015qva,Nomura:2017jxb,Singirala:2017cch,Geng:2017foe,Okada:2018tgy,Das:2018tbd,Das:2017deo,Das:2019fee,Choudhury:2020cpm,Asai:2020xnz}. 
Rather, the so-called type-II seesaw mechanism by a $SU(2)$ triplet Higgs seems to be a simple way of nonvanishing neutrino mass generation~\cite{Schechter:1980gr,Magg:1980ut,Cheng:1980qt}, because a Dirac neutrino mass terms are not necessary there~\cite{Mahapatra:2020dgk,Ghosh:2021khk}.

In this paper we propose the triplet Higgs models in the extra $U(1)_X$ model with the alternative charge for right-handed neutrinos.
The triplet Higgs fields have to be charged under the extra $U(1)_X$ to have Yukawa interactions with lepton doublets
 which are also charged under the $U(1)_X$.
Simultaneously, the triplet Higgs field requires the introduction of another doublet Higgs to develop
 the vacuum expectation value (VEV) by a scalar trilinear term. 
This second Higgs doublet cannot couple with the SM fermions due to the charge mismatch. 
In this construction, the charge assignment requires that the Higgs sector must be the combination of so-called type-I two Higgs doublet model (THDM)~\cite{Branco:2011iw}
 and triplet Higgs model. 
The smallness of the generated neutrino mass via the type-II seesaw mechanism
 is a consequence of the smallness of the second Higgs VEV.
In a philosophical sense, this is a neutrinophilic Higgs model~\cite{Ma:2000cc} in the bosonic sector.
Since the small triplet Higgs VEV is naturally realized in the wide range of parameter space,
 we may expect it easier to observe the Majorana nature through the same sign dilepton signal from the doubly-charged Higgs boson in collider experiments. 
The simplest way to realize the $U(1)_X$ breaking is to introduce an SM singlet scalar with $U(1)_X$ charge $1$ whose VEV generates not only the mass of pseudoscalar but also the mass of right-handed neutrino DM. 
In the minimal extension, the singlet scalar plays three roles. 
This model with only one $U(1)_X$ charged scalar predicts a Dirac right-handed neutrino and one massless right-handed neutrino
 at the renormalizable level.\footnote{For a model with nonrenormalizable terms, see Ref.~\cite{Asai:2020xnz}.}
This could be interesting from the viewpoint of cosmology,
 because those states are candidates of DM and dark radiation (DR), respectively.
The DR has been constrained by cosmological observations~\cite{Planck:2018vyg}, and
 the DR constraint on the $U(1)_{B-L}$ model has been studied~\cite{Heeck:2014zfa,FileviezPerez:2019cyn}.
On the other hand, DR is interesting, because it could relax the so-called
 Hubble tension\footnote{For a review, see for example Ref.~\cite{DiValentino:2021izs}}
 which is the discrepancy between the current Hubble parameter $H_0$ inferred from the cosmic microwave background by Planck~\cite{Planck:2018vyg} and that measured by low-$z$ observations such as the SH0ES collaboration~\cite{Riess:2019cxk}. 
The preferred value of the number of effective neutrino species $N_\mathrm{eff}$ is evaluated as 
$3.2 \lesssim N_\mathrm{eff} \lesssim 3.5$~\cite{Planck:2018vyg}
 and $3.2 \lesssim N_\mathrm{eff} \lesssim 3.4$~\cite{Seto:2021xua,Seto:2021tad} for different data sets.

This paper is organized as follows. 
In Sec.~II, we describe the extra $U(1)_X$ model with the alternative charge assignment for right-handed neutrinos and
 the generation of neutrino masses by the type-II seesaw mechanism. 
The mass spectrum of the particles, especially the various Higgs particles and right-handed neutrinos will be derived.
We also summarize the present experimental constraints on the model.
In Sec.~\ref{sec:cosmology} we provide the relevant formula for discussion of DM and DR.
In Sec.~\ref{sec:plots} we present the interesting parameter region in terms of the current experimental
 and cosmological constraints for the different parameter sets of the $U(1)_X$ model. 
Section~V is devoted to our summary.

\section{Model}

\subsection{$U(1)_X$ model with a triplet scalar}

\begin{table}[h]
	\centering
	\begin{tabular}{|c|ccc|c|} \hline	
		 & $SU(3)_C$ & $SU(2)_L$ & $U(1)_Y$  & $U(1)_X $ \\ \hline
		$q_L^i$ & $\mathbf{3}$ & $\mathbf{2}$ & $\frac{1}{6}$ & $\frac{1}{6}x_H+\frac{1}{3}$ \\
		$u_R^i$ & $\mathbf{3}$ & $\mathbf{1}$ & $\frac{2}{3}$ & $\frac{2}{3}x_H+\frac{1}{3}$ \\
		$d_R^i$ & $\mathbf{3}$ & $\mathbf{1}$ & $-\frac{1}{3}$ & $-\frac{1}{3}x_H+\frac{1}{3}$ \\ \hline
		$l_L^i$ & $\mathbf{1}$ & $\mathbf{2}$ & $-\frac{1}{2}$ & $-\frac{1}{2}x_H-1$ \\
		$e_R^i$ & $\mathbf{1}$ & $\mathbf{1}$ & $-1$ & $-x_H-1$ \\
		$\nu_R^1$ & $\mathbf{1}$ & $\mathbf{1}$ & $0$ & $-4$ \\ 
		$\nu_R^2$ & $\mathbf{1}$ & $\mathbf{1}$ & $0$ & $-4$ \\ 
		$\nu_R^3$ & $\mathbf{1}$ & $\mathbf{1}$ & $0$ & $5$ \\ \hline
            $\Phi_1$ & $\mathbf{1}$ & $\mathbf{2}$ & $\frac{1}{2}$ & $\frac{1}{2}x_H+1$ \\
            $\Phi_2$ & $\mathbf{1}$ & $\mathbf{2}$ & $\frac{1}{2}$ & $\frac{1}{2}x_H$ \\ 
            $\Delta_3$ & $\mathbf{1}$ & $\mathbf{3}$ & $1$ & $x_H+2$ \\ \hline
            $\Phi_X$ & $\mathbf{1}$ & $\mathbf{1}$ & $0$ & $1$ \\ \hline
	\end{tabular}
\caption{
In addition to the SM particle content ($i=1,2,3$), three right-handed neutrinos  
 $\nu_R^i$ ($i=1, 2, 3$), one Higgs doublet $\Phi_1$, one triplet Higgs $\Delta_3$,
 and one $U(1)_X$ Higgs field $\Phi_X$ are introduced. 
$x_H$ is a real free parameter in the $U(1)_X$ charge unfixed by the anomaly-free conditions.
}
\label{table1}
\end{table}

Our model is based on the gauge group $SU(3)_C \times SU(2)_L \times U(1)_Y \times U(1)_{X}$.
The particle content is listed in Table~\ref{table1}.
Under these gauge groups, three generations of right-handed neutrinos ($\nu_R^i$ with $i$ running $1,2,3$)
 have to be introduced for the anomaly cancellation. 
We consider the alternative $U(1)_{B-L}$ charge assignment for right-handed neutrinos.
The $U(1)_X$ symmetry is defiend as the linear combination of $U(1)_Y$ and $U(1)_{B-L}$,
 and the $U(1)_X$ charge is parametrized by the relative $U(1)_Y$ charge $x_H$ normalized
 by the $U(1)_{B-L}$ charge.
For this charge assignment, right-handed neutrinos can not have Yukawa interaction with left-handed leptons $l_L^i$ and the SM Higgs field $\Phi_2$ due to the mismatch of the charge. Instead of introducing two doublet Higgs fields and two singlet Higgs fields to generate Majorana masses of right-handed neutrinos and to form Dirac neutrino masses for the type-I seesaw~\cite{Okada:2018tgy}, we introduce, as an economical way, one triplet Higgs field $\Delta_3$ and one doublet field $\Phi_1$, which can generate neutrino masses by type-II seesaw, and one singlet field $\Phi_X$ to break the $U(1)_X$ symmetry. 
Moreover, the SM singlet scalar with $U(1)_X$ charge $1$ generates not only the mass of pseudoscalar but also the mass of right-handed neutrino DM. 
Thus, this extension of the scalar sector is the minimal and unique. 
The additional $\Phi_1$ does not have any Yukawa coupling to the SM fermion as in the type-I THDM but is necessary to have a trilinear term of $\Delta_3$. 

The Yukawa couplings with those additional Higgs fields are given by
\begin{align}
\mathcal{L}_\mathrm{Yukawa} \supset  -\frac{1}{\sqrt{2}}Y_{\Delta}^{ij}\overline{l_L^{i~C}}\cdot \Delta l_L^j - \sum_{i=1,2} Y_{\nu_R^i} \Phi_X^{\dagger} \overline{\nu_R^{3~C}} \nu_R^i  + \mathrm{ H.c.} ,
\label{Lag1} 
\end{align}
 where the superscript $C$ denotes the charge conjugation, the dot denotes the antisymmetric product of $SU(2)$, $Y_{\Delta}$ and $Y_{\nu_R^i}$ are Yukawa couplings between left-handed lepton doublets and right-handed neutrinos, respectively.
After the triplet Higgs field develops a VEV $v_{\Delta}$, left-handed neutrino masses are generated as
\begin{align}
(\mathcal{M}_\nu)_{ij} =  Y_{\Delta}^{ij} v_{\Delta}.
\label{Eq:neutrino mass}
\end{align}
After $\Phi_X$ develops a VEV $v_X$, the Yukawa interactions between $\nu_R^i$ with $\Phi_X$ give the Dirac mass of right-handed neutrinos as
\begin{align}
\mathcal{L}_{\nu_R \mathrm{mass}} = &
 - ( \nu_R^1{}^{\dagger} \,\,  \nu_R^2{}^{\dagger}  \,\, \nu_R^3{}^T  )
\left(
\begin{array}{ccc}
0 & 0 & Y_{\nu_R^1} \\
0 & 0 & Y_{\nu_R^2} \\
Y_{\nu_R^1} & Y_{\nu_R^2} & 0 \\
\end{array}
\right)
\frac{v_X}{\sqrt{2}}
\left(
\begin{array}{c}
 \nu_R^1{} \\
 \nu_R^2{} \\
 \nu_R^3{}^* \\
\end{array}
\right) \nonumber \\ 
  = & - ( \nu_R^1{}^{\dagger} \,\,  \nu_R^2{}^{\dagger}  \,\, \nu_R^3{}^T  ) U^{\dagger}
\left(
\begin{array}{ccc}
0 & 0 & 0 \\
0 & 0 & m \\
0 & m & 0 \\
\end{array} 
\right) U 
\left(
\begin{array}{c}
 \nu_R^1{} \\
 \nu_R^2{} \\
 \nu_R^3{}^* \\
\end{array}
\right) , 
\label{eq:nuRmass}
\end{align}
 where we define
\begin{align}
U= &\left(
\begin{array}{ccc}
 \frac{m_2}{m} & -\frac{m_1}{m} & 0 \\
 \frac{m_1}{m} & \frac{m_2}{m} & 0 \\
 0 & 0 & 1 \\
\end{array}
\right) , \\
m_1 = & Y_{\nu_R^1}\frac{v_X}{\sqrt{2}} , \\
m_2 = & Y_{\nu_R^2}\frac{v_X}{\sqrt{2}} , \\
m =& \sqrt{m_1^2+m_2^2} =  \sqrt{\frac{Y_{\nu_R^1}^2+Y_{\nu_R^2}^2}{2}} v_X .
\end{align}
We find that one linear combination
\begin{align}
\nu_R =  \frac{m_2}{m}\nu_R^1-\frac{m_1}{m}\nu_R^2, 
\end{align}
 is a massless state.
Other two components are summarized as
\begin{align}
\mathcal{L}_{\nu_R \mathrm{mass}} = & 
 - \left( \frac{m_1}{m}\nu_R^1{}^{\dagger}+\frac{m_2}{m}\nu_R^2{}^{\dagger}  \,\,\, \nu_R^3{}^T  \right)
\left(
\begin{array}{cc}
0 & 1  \\
1 & 0  \\
\end{array}
\right)
m
\left(
\begin{array}{c}
 \frac{m_1}{m}\nu_R^1+\frac{m_2}{m}\nu_R^2 \\
 \nu_R^3{}^* \\
\end{array}
\right)  \nonumber \\ 
 = & - \overline{\chi}m \chi,
\end{align}
by composing a Dirac spinor as
\begin{align}
\chi = & \left(
\begin{array}{c}
 \frac{m_1}{m}\nu_R^1+\frac{m_2}{m}\nu_R^2 \\
 \nu_R^3{}^* \\
\end{array}
\right) .
\end{align}

There is one massless state $\nu_R$ and one Dirac fermion $\chi$.
Thus, $\nu_R$ behaves as DR and $\chi$ is a candidate for DM.
It is worth noting that $\chi$ has no direct coupling with the SM particles thanks to its $U(1)_X$ charge assignment which guarantees the stability of the DM candidate. 
This is in a remarkable contrast with right-handed neutrino DM 
 in the minimal $U(1)_X$ model with the standard charge assignment,
 where the extra $Z_2$ parity has to be introduced by hand
 to stabilize DM~\cite{Okada:2010wd,Okada:2016gsh,Okada:2016tci}.\footnote{For a review on this class of models, see e.g., Ref.~\cite{Okada:2018ktp}.}

The gauge interactions of $\nu_R$ and $\chi$ can be read from those of $\nu_R^i$
\begin{align}
\mathcal{L}_\mathrm{int} = & \overline{\nu_R^i}i\gamma^{\mu}\left(\partial_{\mu} - i q_{\nu_R^i} g_X X_{\mu}\right)\nu_R^i  \nonumber \\
= & i \overline{\nu_R} \gamma^{\mu}  \left( \partial_{\mu}- i (-4)g_X X_{\mu} \right)\nu_R
+ i \overline{\chi} \left( \gamma^{\mu} \partial_{\mu}- i ( (-4)P_R +(-5) P_L ) g_X X_{\mu} \right) \chi ,
\end{align}
while similarly, the $U(1)_X$ gauge interaction for an SM chiral fermion ($f_{L/R}$) can be read from the usual covariant derivative, 
\begin{align}
\mathcal{L}_\mathrm{int} = \sum_{f_{L/R}} i \overline{f_{L/R}}\gamma^{\mu}\left( - i q_{f_{L/R}} \right)  g_X X_{\mu}f_{L/R} ,
\end{align}
 where $q_{f_{L/R}}$ is a $U(1)_X$ charge of $f_{L/R}$ listed in Table~\ref{table1}.

The scalar potential is given by
\begin{align}
V =& V_1 +V_2, \label{eq:V_UV} \\
V_1 =& +\hat{\mu_1}^2 |\Phi_1|^2 -\hat{\mu_2}^2 |\Phi_2|^2 \nonumber\\ 
 &  + \frac{1}{2}\lambda_1|\Phi_1|^4   +
 \frac{1}{2}\lambda_2|\Phi_2|^4 + \lambda_3|\Phi_1|^2|\Phi_2|^2 +\lambda_4 |\Phi_1^\dagger \Phi_2|^2 \nonumber\\ 
 & + \hat{\mu_3}^2 \mathrm{Tr}(\Delta_3^{\dagger}\Delta_3) + \frac{1}{2}\Lambda_1 \left(\mathrm{Tr}(\Delta_3^{\dagger}\Delta_3)\right)^2
 + \frac{1}{2}\Lambda_2 \left( \left(\mathrm{Tr}(\Delta_3^{\dagger}\Delta_3)\right)^2- \mathrm{Tr}[(\Delta_3^{\dagger}\Delta_3)^2] \right) \nonumber\\ 
 &  + \left(\Lambda_{41}|\Phi_1|^2+\Lambda_{42}|\Phi_2|^2 \right) \mathrm{Tr}(\Delta_3^{\dagger}\Delta_3) +
 \Lambda_{51}\Phi_1^{\dagger}[\Delta_3^{\dagger},\Delta_3]\Phi_1 + \Lambda_{52}\Phi_2^{\dagger}[\Delta_3^{\dagger},\Delta_3]\Phi_2 \nonumber\\ 
 &  - \frac{\Lambda_6}{\sqrt{2}} (\Phi_1^T \cdot \Delta_3 \Phi_1 + \mathrm{ H.c.} ) ,\\
V_2 =& -\mu_X^2 |\Phi_X|^2 + \frac{1}{2}\lambda_X |\Phi_X|^4 + \left( \lambda_{12} \Phi_X (\Phi_1^\dagger \Phi_2) + \mathrm{ H.c.} \right) \nonumber\\ 
 &  + \lambda_{X1}|\Phi_X|^2|\Phi_1|^2 +\lambda_{X2}|\Phi_X|^2|\Phi_2|^2 + \lambda_{X\Delta}|\Phi_X|^2 \mathrm{Tr}(\Delta_3^{\dagger}\Delta_3) . \label{eq:V2} 
\end{align}
At a high scale $\gg v \simeq 246$ GeV, $U(1)_X$ symmetry is spontaneously broken by the VEV of $\Phi_X = v_X/\sqrt{2} =\mu_X^2/\lambda_X$. 
At the $U(1)_X$ broken vacuum, the $U(1)_X$ gauge boson acquires the mass,
\begin{align}
m_X^2 = g_X^2 v_X^2,
\end{align}
 as $\nu_R$ does in Eq.~(\ref{eq:nuRmass}).
Then, the scalar $\phi_X$ from $\Phi_X$ also has the mass $m_{\phi_X}^2=\lambda_{X}v_X^2$.
Thus, the effective scalar potential at a low-energy scale below the scale of $v_X$ is given by
\begin{align}
V  =& +\mu_1^2 |\Phi_1|^2 -\mu_2^2 |\Phi_2|^2 - \left( \mu_{12}^2 (\Phi_1^\dagger \Phi_2) + \mathrm{ H.c.} \right) \nonumber\\ 
 &  + \frac{1}{2}\lambda_1|\Phi_1|^4   +
 \frac{1}{2}\lambda_2|\Phi_2|^4 + \lambda_3|\Phi_1|^2|\Phi_2|^2 +\lambda_4 |\Phi_1^\dagger \Phi_2|^2 \nonumber\\ 
 & + \mu_3^2 \mathrm{Tr}(\Delta_3^{\dagger}\Delta_3) + \frac{1}{2}\Lambda_1 \left(\mathrm{Tr}(\Delta_3^{\dagger}\Delta_3)\right)^2
 + \frac{1}{2}\Lambda_2 \left( \left(\mathrm{Tr}(\Delta_3^{\dagger}\Delta_3)\right)^2- \mathrm{Tr}[(\Delta_3^{\dagger}\Delta_3)^2] \right) \nonumber\\ 
 &  + \left(\Lambda_{41}|\Phi_1|^2+\Lambda_{42}|\Phi_2|^2 \right) \mathrm{Tr}(\Delta_3^{\dagger}\Delta_3) +
 \Lambda_{51}\Phi_1^{\dagger}[\Delta_3^{\dagger},\Delta_3]\Phi_1 + \Lambda_{52}\Phi_2^{\dagger}[\Delta_3^{\dagger},\Delta_3]\Phi_2 \nonumber\\ 
 &  - \frac{\Lambda_6}{\sqrt{2}} (\Phi_1^T \cdot \Delta_3 \Phi_1 + \mathrm{ H.c.} ) ,
\label{eq:V}
\end{align}
 where the third term is generated from the third term in Eq.~(\ref{eq:V2}) by replacing $\Phi_X$
 with its VEV as $-\mu_{12}^2 = \lambda_{12}v_X/\sqrt{2} $.
$\mu_i^2 = \hat{\mu_i}^2 + \lambda_{iX}v_X^2/2$, $\lambda_i$, and $\Lambda_{i(j)}$ are coupling constants.
Unlike the usual THDM, the $(\Phi_1^\dagger \Phi_2)^2$ term is absent due to the $U(1)_X$ gauge symmetry.
The $\mu_{12}^2$ terms generated by the VEV of $v_X$ is essential
 to give the mass of a $CP$ odd Higgs boson and remove a dangerous Nambu-Goldstone boson from the spectrum. 

The stationary conditions are expressed as
\begin{align}
\mu_1^2 &= \frac{2 \mu_{12}^2 v_2-v_1 \left(\lambda_1 v_1^2+v_2^2 (\lambda_3+\lambda_4)+v_{\Delta}^2 (\Lambda_{41}-\Lambda_{51})-2 \Lambda_6 v_{\Delta}\right)}{2 v_1},\\
\mu_2^2 &= \frac{\lambda_2 v_2^3+v_1^2 v_2 (\lambda_3+\lambda_4)+v_2 v_{\Delta}^2 (\Lambda_{42}-\Lambda_{52})-2 \mu_{12}^2 v_1}{2 v_2},\\ 
\mu_3^2 &= \frac{\Lambda_6 v_1^2-v_{\Delta} \left(\lambda_1 v_{\Delta}^2+v_1^2 (\Lambda_{41}-\Lambda_{51})+v_2^2 (\Lambda_{42}-\Lambda_{52})\right)}{2 v_{\Delta}}, \label{eq:stat.3}
\end{align}
where $v_1$ and $v_2$ are the VEVs of $\Phi_1$ and $\Phi_2$, respectively.
Our notation satisfies $v=\sqrt{v_1^2+v_2^2} \simeq 246$ GeV.
The condition (\ref{eq:stat.3}) can be recast as
\begin{align}
\Lambda_6  = \frac{v_{\Delta} \left(\lambda_1 v_{\Delta}^2+v_1^2 (\Lambda_{41}-\Lambda_{51})+v_2^2 (\Lambda_{42}-\Lambda_{52})+2 \mu_3^2\right)}{v_1^2} \simeq \frac{v_{\Delta} 2\mu_3^2 }{v_1^2},
\end{align}
 where, in the last approximation, we have assumed the condition for the type-II seesaw mechanism, $\mu_3^2 \gg v^2$. For a given $\Lambda_6$ and $\mu_3^2$, our smaller $v_1$ results a smaller $v_{\Delta}$, which naturally fits the $\rho$ parameter constraint and enhances the decay rate of the doubly-charged Higgs boson to the same sign dilepton $H^{\pm\pm} \rightarrow \ell_i^{\pm} \ell_j^{\pm}$ as we will show. 
In the small $v_{\Delta}$ limit, the masses of scalars are given by
\begin{align}
m_{h/H}^2 =& \left(
\begin{array}{cc}
 \lambda_1 v_1^2+\mu_{12}^2\frac{v_2}{v_1} & v_1 v_2 (\lambda_3+\lambda_4)-\mu_{12}^2 \\
 v_1 v_2 (\lambda_3+\lambda_4)-\mu_{12}^2 & \lambda_2 v_2^2+\mu_{12}^2\frac{v_1}{v_2} \\
\end{array}
\right)  , \\
m_{H_2}^2 =& m_{H_2^{\pm}}^2 -\frac{1}{2} \left(\Lambda_{51} v_1^2+\Lambda_{52} v_2^2\right), \\
m_{A_1}^2 =& \mu_{12}^2 \left(\frac{v_1}{v_2}+\frac{v_2}{v_1}\right), \\
m_{A_2}^2 =& m_{H_2}^2, \\
m_{H_1^{\pm}}^2 =& m_{A_1}^2-\frac{\lambda_4 }{2}v^2 , \\
m_{H_2^{\pm}}^2 =& \frac{1}{2} \left(\Lambda_{41} v_1^2+\Lambda_{42} v_2^2+2 \mu_3^2\right), \\
m_{H^{\pm\pm}}^2 =& m_{H_2^{\pm}}^2 +\frac{1}{2} \left( \Lambda_{51}v_1^2 + \Lambda_{52}v_2^2 \right) . 
\end{align}

\subsection{Summary of experimental constraints}

\subsubsection{$\rho$ parameter}

The $\rho$ parameter in this model is given as
\begin{align}
 \rho \equiv \frac{m_W^2}{m_Z^2 c_W^2} = \frac{1+\frac{2 v_{\Delta}^2}{v^2}}{1+\frac{4 v_{\Delta}^2}{v^2}},
\end{align}
which is experimentally constrained as $\rho = 1.00038 \pm 0.00020 $~\cite{Zyla:2020zbs}.
We find $v_{\Delta} \lesssim 0.78  (2.6) $ GeV at the $2 (3)$ sigma.

\subsubsection{The Z boson decay width}

Nonobservation of the exotic decay of the $Z$ boson constrains the mass of the doubly-charged Higgs boson as $m_Z < 2 m_{H^{\pm\pm}}$~\cite{Kanemura:2013vxa}.

\subsubsection{Doubly-charged Higgs boson search at the LHC}

The LHC bound on the doubly-charged Higgs boson has been derived as
 $m_{H^{\pm\pm}} \gtrsim 880$ GeV for $H^{\pm\pm} \rightarrow \ell^{\pm}\ell^{\pm}$~\cite{CMS:2012dun,ATLAS:2017xqs} and 
 $m_{H^{\pm\pm}} \gtrsim 350$ GeV for $H^{\pm\pm} \rightarrow W^{\pm}W^{\pm} $~\cite{ATLAS:2018ceg,ATLAS:2021jol}.

\subsubsection{Lepton flavor violation}

In a triplet Higgs model, lepton flavor-violating decays of a charged lepton are induced at tree level~\cite{Chun:2003ej,Kakizaki:2003jk,Akeroyd:2009nu}.
The branching ratio is given by~\cite{Kakizaki:2003jk}
\begin{equation}
\mathrm{Br}(\ell_i \rightarrow \bar{\ell}_j \ell_k \ell_l) = \frac{1}{64 G_F^2 m^4_{H^{\pm\pm}}}
 \left| \frac{(Y_{\Delta}^{ij})^{\dagger}Y_{\Delta}^{kl}}{2} \right|^2 .
\end{equation}
The stringent bound is $ \mathrm{Br}(\mu \rightarrow \bar{e} e e) < 1.0 \times 10^{-12} $ from the SINDRUM experiment~\cite{SINDRUM:1987nra}.

A flavor-violating radiative decay $\ell_i \rightarrow \ell_j \gamma$ is also induced and
 its branching ratio for $\mu \rightarrow e \gamma$, which gives the most stringent bound, is given by~\cite{Kakizaki:2003jk}
\begin{equation}
\mathrm{Br}(\mu \rightarrow e \gamma) = \frac{48 \pi^3 \alpha_\mathrm{em}}{G_F^2 m^4_{H^{\pm\pm}}}
 \left| (Y_{\Delta}^{\dagger}Y_{\Delta})_{e \mu}\frac{1}{16 \pi^2}\frac{3}{16} \right|^2 .
\end{equation}
The MEG experiment has reported the latest result of $\mathrm{Br}(\mu \rightarrow e \gamma) < 4.2 \times 10^{-13}$~\cite{MEG:2016leq}.

Any of those lepton flavor-violating decay gives the lower bound on the VEV of the triplet Higgs field as $v_{\Delta} > \mathcal{O}(1)$ eV~\cite{Antusch:2018svb} for $m_{H^{\pm\pm}} \lesssim 1$ TeV.

\subsubsection{$X$ boson search at the LHC}

In this section we evaluate the production cross section of this process at
 the LHC for a choice of parameters consistent with the LHC constraints 
 from dilepton channel $pp \rightarrow X \rightarrow \ell\bar{\ell}$~\cite{CMS:2018ipm,ATLAS:2019erb} and the dijet constraints~\cite{CMS:2018wxx,ATLAS:2019bov}.
The decay rates of $X$ are given by
\begin{align}
\sum_{f=\mathrm{quarks, leptons}}\Gamma_X(X\rightarrow f\bar{f}) &= \frac{g_X^2}{24 \pi}m_X F(x_H), \label{Eq:GammaXff}\\
\Gamma_X(X\rightarrow \nu_R\overline{\nu_R}) &= \frac{g_X^2}{24 \pi}m_X 16 , \label{Eq:GammaXnunu}\\ 
\Gamma_X(X\rightarrow \chi\overline{\chi}) &= \frac{g_X^2}{24 \pi} \sqrt{m_X^2-4m_{\chi}^2} \left( 41m_X^2 -120 m_{\chi}^2 \right), \label{Eq:GammaXchichi}
\end{align}
with the auxiliary function~\cite{Okada:2016tci}
\begin{equation} 
 F(x_H) = 13 + 16 x_H + 10 x_H^2 .
\end{equation}
$g_X$ has been constrained to be small in previous studies.
As in Ref.~\cite{Das:2019fee}, since the total $X$ boson decay width is very narrow, we use the narrow width approximation to evaluate the $X$ boson production cross section 
\begin{align}
\sigma(pp \rightarrow X) &= 2\sum_{q,\bar{q}}\int dx\int dy f_q(x,Q) f_{\bar{q}}(x,Q)\hat{\sigma}(\hat{s}) , \label{Eq:sigma_ppX}\\
\hat{\sigma}(\hat{s}) &= \frac{4\pi^2}{3}\frac{\Gamma_X(X \rightarrow q\bar{q})}{m_X}\delta(\hat{s}-m_X^2) , 
\end{align}
where $f_q$ and $f_{\bar{q}}$ are the parton distribution function (PDF) for a quark and antiquark, $\hat{s}=x y s$ is the invariant mass squared of colliding quarks for the center of mass energy $s$.
The factor $2$ in Eq.~(\ref{Eq:sigma_ppX}) counts two ways of $q$ coming from which proton out of two colliding protons.
Since the most severe bound is from the dilepton channel ($\ell=e, \mu$), we calculate $\sigma(pp\rightarrow X) \mathrm{Br}(X\rightarrow \ell\bar{\ell})$ with
\begin{equation}
\mathrm{Br}(X\rightarrow \ell\bar{\ell}) = \frac{8+12x_H+5x_H^2}{4 F(x_H)} ,
\end{equation}
and compare it with the ATLAS results~\cite{ATLAS:2019erb}.
We employ PDFs of CTEQ6L~\cite{Pumplin:2002vw} with a factorization scale $Q = m_X$ for simplicity.
Following the manner to obtain a suitable $k$-factor presented in Ref.~\cite{Okada:2016gsh}, we scale our result by a $k$-factor of $k = 0.947$ to match the recent ATLAS analysis in our calculation. 
The results will be presented after we discuss cosmological constraints.

\subsection{Decay of doubly-charged Higgs boson}

The decay rates for principal decay modes of $H^{\pm\pm}$ are~\cite{Aoki:2011pz,BhupalDev:2013xol}
\begin{align}
\Gamma( H^{\pm\pm} \rightarrow \ell_i^{\pm} \ell_j^{\pm}) & \simeq 
  S_{ij}\frac{|Y_{\Delta}^{ij}|^2}{2}\frac{m_{H^{\pm\pm}}}{4\pi} , \label{Eq:Brellell} \\
\Gamma( H^{\pm\pm}\rightarrow W^{\pm} W^{\pm}) & \simeq  \frac{ g^4 v_{\Delta}^2 m^3_{H^{\pm\pm}}}{64\pi m_W^4}  \left( \frac{3m_W^4}{m_{H^{\pm\pm}}^4} -\frac{m_W^2}{m_{H^{\pm\pm}}^2} +\frac{1}{4}\right) ,
\end{align}
with
\begin{align}
 S_{ij} = \left\{ 
\begin{array}{c}
 1 \\
 \frac{1}{2} \\
\end{array} 
\qquad \mathrm{for} \quad
\begin{array}{c}
 i \neq j \\
 i = j \\
\end{array} 
\right. .
\end{align}
Those of other minor modes, which are not relevant for our later discussion, can be found in Ref.~\cite{Aoki:2011pz}.

\begin{figure}[thb]
	\begin{center}
		\begin{tabular}{cc}
			\includegraphics[width=0.5\textwidth]{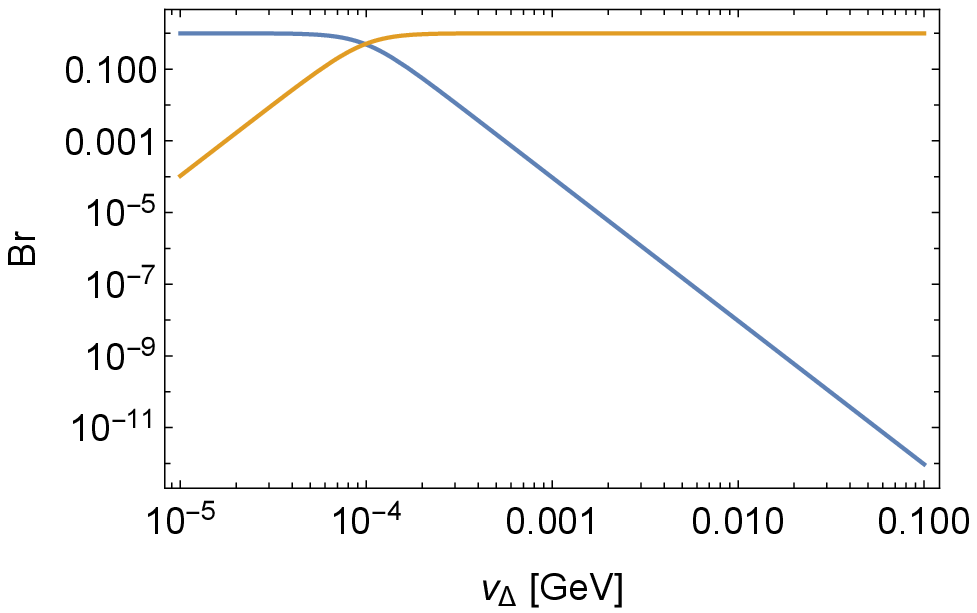}
			 &
			\includegraphics[width=0.5\textwidth]{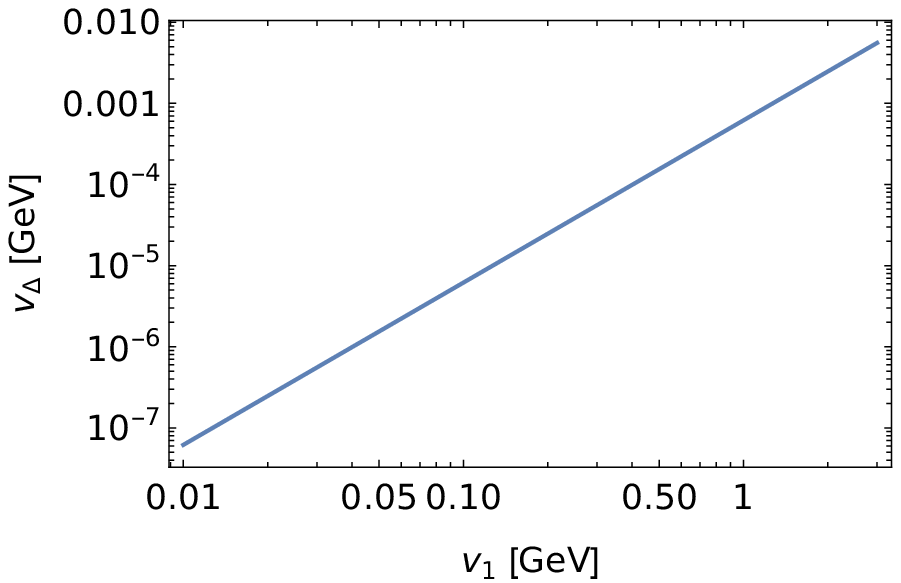}
		\end{tabular}
	\end{center}
\caption{\textit{Left}: The decay branching ratio of $H^{\pm\pm}$ for $m_{H^{\pm\pm}}=900$ GeV. The light blue and orange curves indicate $\mathrm{Br}(H^{\pm\pm} \rightarrow \ell^{\pm}\ell^{\pm})$ and $\mathrm{Br}(H^{\pm\pm} \rightarrow W^{\pm}W^{\pm})$, respectively. $v_{\Delta} \simeq 10^{-4}$ GeV is the critical  value at which the dominant mode changes.  
\textit{Right}: The $v_1$ dependence of $v_{\Delta}$. This is for $\Lambda_6 = 10^3$ GeV and $m_{H^{\pm\pm}}=900$ GeV. A very small $v_{\Delta}$ can be easily achieved with a sub-GeV scale $v_1$. }
\label{Fig:Br}
\end{figure}

In the left panel of Fig.~\ref{Fig:Br}, we show the decay branching ratio $\mathrm{Br}(H^{\pm\pm} \rightarrow \ell^{\pm}\ell^{\pm})$ with the light blue curve and $\mathrm{Br}(H^{\pm\pm} \rightarrow W^{\pm}W^{\pm})$ with the orange curve. Here, we have taken $m_{H^{\pm\pm}}=900$ GeV and substituted the formula of neutrino mass (\ref{Eq:neutrino mass}) into (\ref{Eq:Brellell}). Then, the magnitude of neutrino masses is taken to be the scale of atmospheric neutrino mass difference of $\mathcal{O}(0.1)$ eV. For $v_{\Delta} \gtrsim 10^{-4}$ GeV, reflecting the Majorana nature of neutrino mass generated by the type-II seesaw mechanism, the lepton number-violating mode $H^{\pm\pm} \rightarrow \ell^{\pm}\ell^{\pm}$ is dominant. The right panel shows that, in our model, a small $v_{\Delta}$ can be realized by a sub-GeV scale $v_1$. This is a distinctive feature of our model. While the tiny $v_{\Delta}$ and, as the result, the smallness of neutrino mass comes from the small dimensionful parameter $\Lambda_6$ in the minimal Higgs triplet model, the smallness of neutrino mass is a result of the smallness of $v_1$ and an energy scale of $\Lambda_6$ being much smaller than the EW scale is not necessarily required. In this sense, our $\Phi_1$ plays the same role as the so-called neutrinophilic Higgs field only in the scalar sector.

\section{Cosmology}
\label{sec:cosmology}

Next, we consider cosmological constraints and implication of our model.

\subsection{Dark radiation}

With one $U(1)_X$ charged scalar $\Phi_X$, out of three right-handed neutrinos, there is one massless state $\nu_R$ and one Dirac fermion $\chi$. We at first consider the DR constraint due to thermal production of $\nu_R$. 
 
The thermal averaged cross section of $\nu_R$ for the temperature $T \ll m_X$ is expressed as
\begin{equation}
\langle \sigma v \rangle \simeq \sum_f \langle \sigma v(f\bar{f}\leftrightarrow \nu_R\nu_R) \rangle ,
\end{equation}
 with
\begin{align}
\langle \sigma v(u\bar{u}\leftrightarrow \nu_R\nu_R) \rangle &= \frac{2 g^4 N_c T^2 (x_H(17 x_H+20)+8)}{9 \pi m_X^4}, \\
\langle \sigma v(d\bar{d}\leftrightarrow \nu_R\nu_R) \rangle &= \frac{2 g^4 N_c T^2 (x_H(5 x_H-4)+8)}{9 \pi m_X^4}, \\
\langle \sigma v(\ell\bar{\ell}\leftrightarrow \nu_R\nu_R) \rangle &= \frac{2 g^4 T^2 (x_H(5 x_H+12)+8)}{\pi m_X^4}, \\
\langle \sigma v(\nu\bar{\nu}\leftrightarrow \nu_R\nu_R) \rangle &= \frac{2 g^4 T^2 (x_H+2)^2}{\pi m_X^4}, 
\end{align}
and the color factor $N_c=3$ for quarks. The invariant squared amplitude before taking thermal averaging are listed in Appendix~\ref{sec:AppA}.
The decoupling temperature $T_\mathrm{dec}$ of $\nu_R$ from the thermal bath is evaluated by
\begin{equation}
 \left. \langle \sigma v \rangle n_{\nu_R}\right|_{T=T_\mathrm{dec}} = H(T_\mathrm{dec}),
\end{equation}
 where $H = \dot{a}/a $ is the cosmic expansion rate in the radiation dominated universe described by
\begin{align}
H^2 &= \frac{1}{3M_P^2}\rho_r , \\
\rho_r &= \frac{\pi^2 g_{*}}{30}T^4 ,
\end{align}
with $a$, $\rho_r$, and $g_{*}$ are the scale factor, the energy density of radiation and the number of relativistic degrees of freedom, respectively. $M_P \simeq 2.4\times 10^{18}$ is the reduced Planck mass.
After $\nu_R$ decoupled from thermal bath at the decoupling temperature $T_\mathrm{dec}$, the energy density of $\nu_R$, $\rho_{\nu_R}$, decreases as $\rho_{\nu_R} \propto a^{-4}$.
By parametrizing the energy density of $\nu_R$ with $\Delta N_\mathrm{eff}$ as
\begin{equation}
\rho_{\nu_R} = \Delta N_\mathrm{eff} \frac{7}{4}\frac{\pi^2}{30} T_{\nu}^4 ,
\end{equation}
the total radiation energy density except photons is expressed as
\begin{equation}
N_\mathrm{eff} = N_\mathrm{eff}^\mathrm{\nu} + \Delta N_\mathrm{eff} ,
\end{equation}
where $T_{\nu}$ is the temperature of left-handed SM neutrinos and $N_\mathrm{eff}^\mathrm{\nu}$ is the effective number of neutrinos in the SM. For recent calculations of $N_\mathrm{eff}^\mathrm{\nu}$, see e.g., Refs.~\cite{Escudero:2018mvt,Bennett:2019ewm,Escudero:2020dfa,Akita:2020szl,Bennett:2020zkv}.

\subsection{Dark matter}

\subsubsection{Abundance}

We estimate the thermal relic abundance of our Dirac DM $\chi$ by solving the Boltzmann equation,
\begin{equation}
 \frac{d n }{dt}+3H n =-\langle\sigma v\rangle ( n^2 - n_{\rm EQ}^2),
\label{eq:boltzman}
\end{equation}
 where $n$ is the number density of $\chi$, $n_{\rm EQ}$ is its number density at thermal equilibrium, $\langle\sigma v\rangle$ is the thermal averaged products of the annihilation cross section and the relative velocity. The amplitude squared integrated over the scattering angle $\theta$ is given by  
\begin{equation}
 \sum_{i = f, \nu_R} \int \overline{|\mathcal{M}(i\bar{i}\leftrightarrow \chi\chi)|^2} d\cos\theta  = \frac{2 g_X^4}{3}\frac{ s (F(x_H) + 16) \left( 79 m_N^2+41 s \right)}{ \left( m_X^2-s \right)^2 +\Gamma_X^2 m_X^2  } 
\end{equation}
 where $s$ is the center-of-mass energy and the total decay width is given by
\begin{equation}
 \Gamma_X = \sum_{i = f,\nu_R,\chi}\Gamma_X(X \rightarrow i\bar{i}) ,
\end{equation}
 with each partial decay width (\ref{Eq:GammaXff}), (\ref{Eq:GammaXnunu}) and (\ref{Eq:GammaXchichi}).

The resultant DM relic abundance is given by
\begin{align}
\Omega_{\chi}h^2 = \frac{1.1 \times 10^9 x_d \mathrm{GeV}^{-1}}{\sqrt{ 8\pi g_*}M_P \langle\sigma v\rangle },
\end{align}
where $x_d = m_{\chi}/T_d$ with the decoupling temperature $T_d$~\cite{Kolb:1990vq}.

\subsubsection{Direct DM detection bound}
\label{subsec:direct_bound}

The DM $\chi$ with the mass $m_{\chi}$ can scatter off nucleons through the
$X$ boson exchange. 
The spin-independent (SI) weakly interacting massive particle (WIMP)-nucleon cross section
 for the elastic scattering through vector-vector couplings is given by~\cite{Jungman:1995df}
\begin{equation}
\sigma_{SI} = \frac{\mu_N^2}{4 \pi}\left( Z b_p+ (A-Z) b_n\right)^2 ,
\label{Eq:directCrossSection}
\end{equation}
 with 
\begin{align}
\mu_N &= \frac{m_\chi m_N}{m_\chi+m_N}, \\
b_p &= \frac{g_\chi}{m_X^2}(2 g_u +g_d), \\
b_n &= \frac{g_\chi}{m_X^2}(g_u +2 g_d), \\
g_\chi &= 4 g_X, \\
g_u &= \left( \frac{1}{2}\left(\frac{x_H}{6}+\frac{2 x_H}{3}\right) +\frac{1}{3} \right)g_X , \\
g_d &= \left( \frac{1}{2}\left(\frac{x_H}{6}-\frac{x_H}{3}\right) +\frac{1}{3} \right)g_X ,
\end{align}
Here, $Z$ and $A$ are the atomic number and the mass number of a target. $m_N$ is the mass of nucleus.

\section{Benchmark points}
\label{sec:plots}

\begin{figure}[thb]
	\centering
		\begin{tabular}{ccc}
			\includegraphics[width=0.5\textwidth]{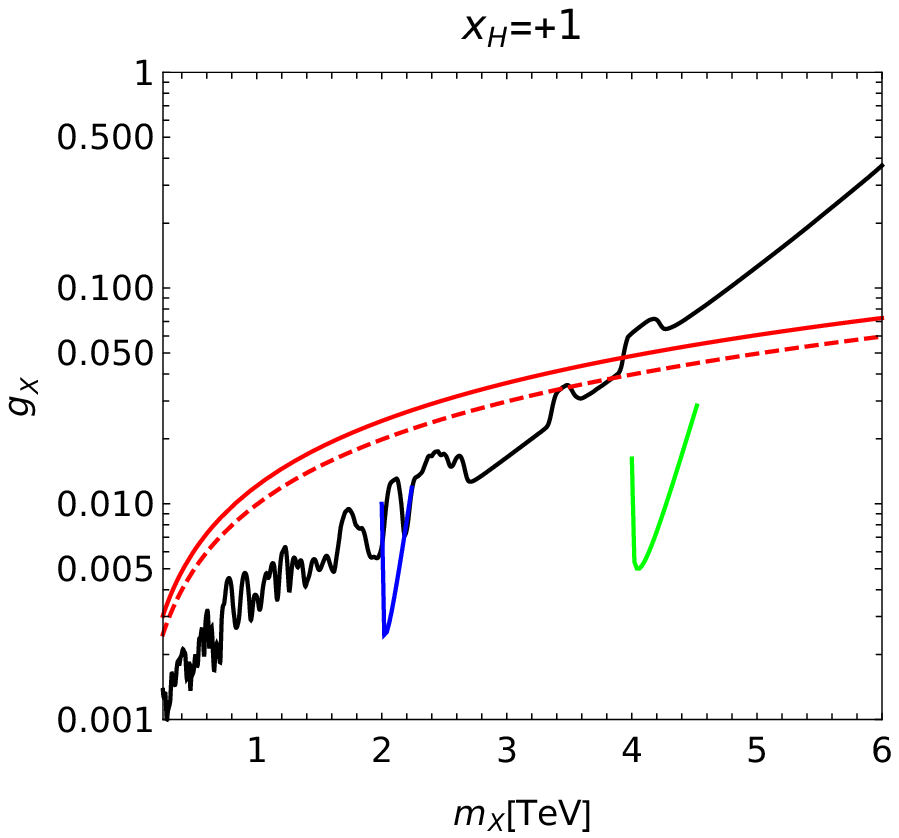}
			 & \qquad &
			\includegraphics[width=0.5\textwidth]{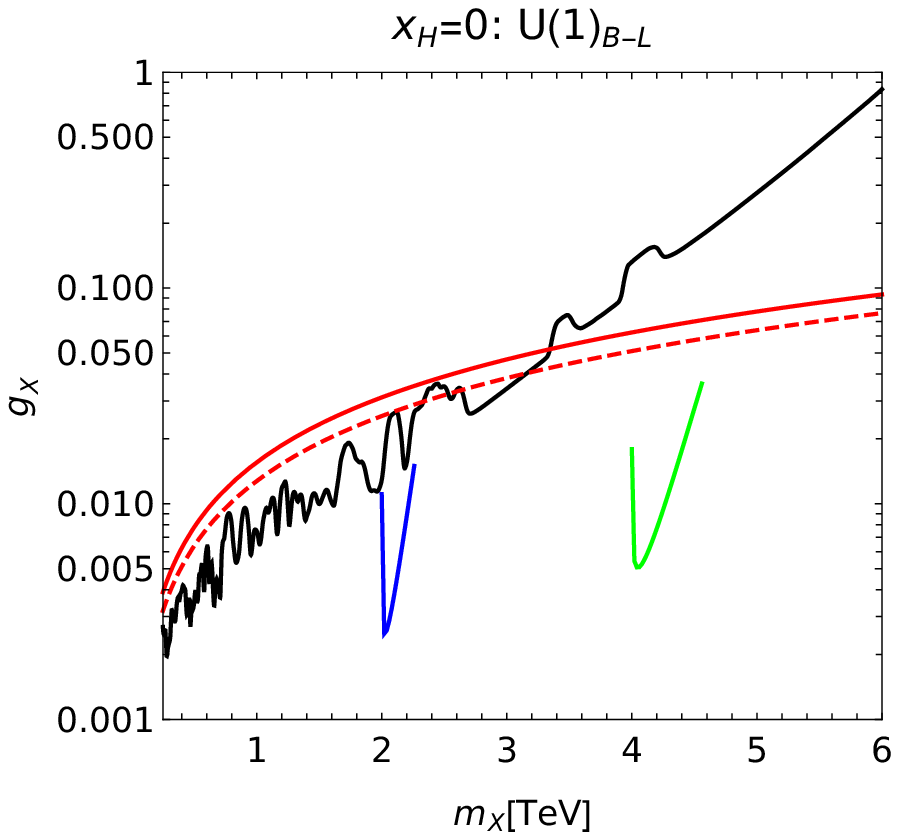} \\[10pt]
			\includegraphics[width=0.5\textwidth]{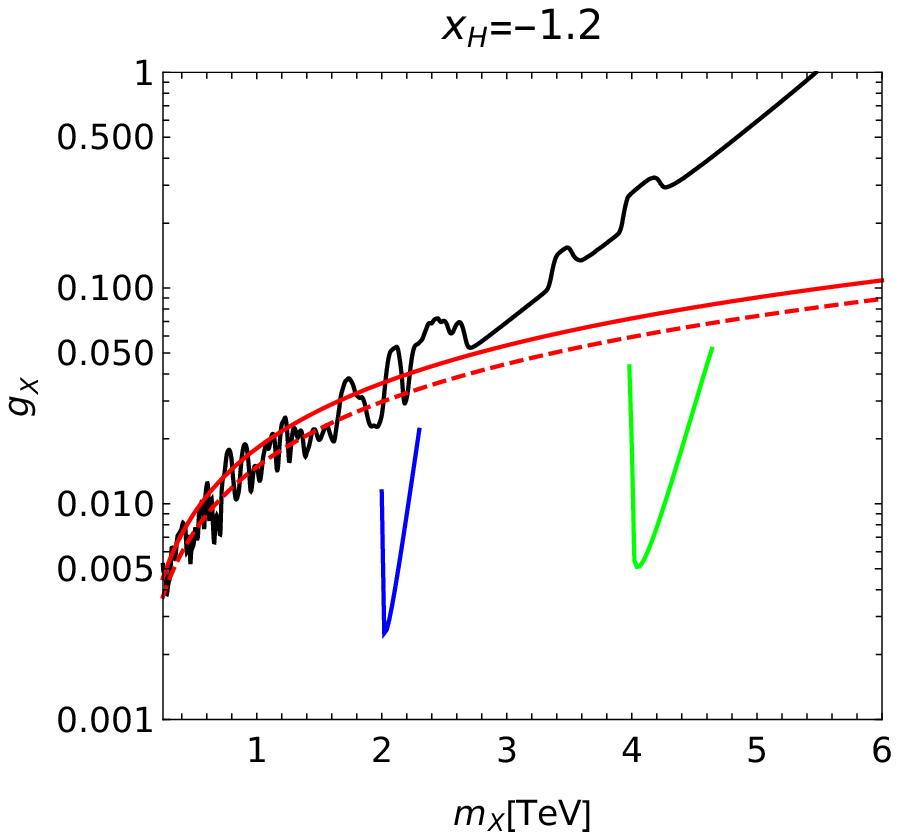}
			 & \qquad &
			\includegraphics[width=0.5\textwidth]{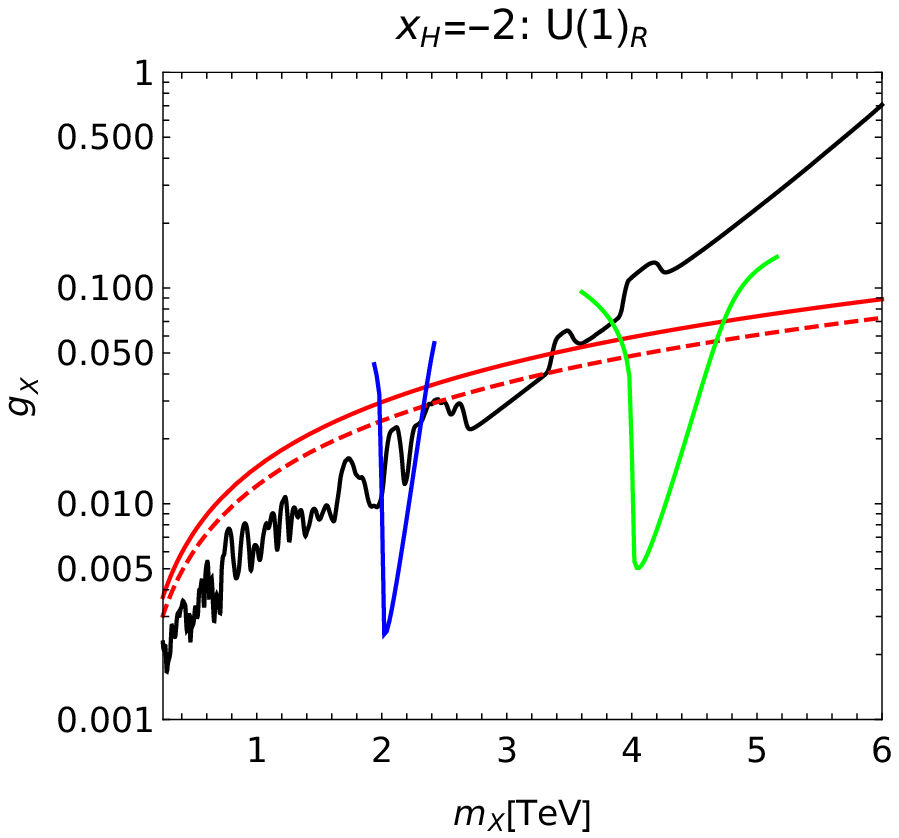}
		\end{tabular}
\caption{Constraints summary. 
LHC(ATLAS) constraints (black), $N_\mathrm{eff} = 3.5 \, (3.1)$ with red solid (dashed) curve, points satisfying both $\Omega_{\chi} h^2 \simeq 0.1$ and the constraints from DM direct searches for $m_N=1$ TeV (blue) and $2$ TeV (green). 
\textit{Upper left}: For the $x_H=1$ case.
\textit{Upper right}: For the $x_H=0; U(1)_{B-L}$ case.
\textit{Lower left}: For the $x_H=-1.2$ case.
\textit{Lower right}: For the  $x_H=-2; U(1)_R$ case.
}
\label{Fig:Constr}
\end{figure}

We examine the constraints on our model for several parameters by collider experiments and cosmology.

The first case is $x_H =0$, which corresponds to $U(1)_{B-L}$. 
On the upper-right panel in Fig.~\ref{Fig:Constr},
 we show the LHC constraints drawn by black solid curve in a $(m_X, g_X)$ plane
 and parameter points where desired thermal DM abundance $\Omega_{\chi}h^2\simeq 0.1$ is reproduced without confronting the latest DM direct search XENON1T(2018)~\cite{XENON:2018voc}. 
The blue and green curves correspond to fixed DM masses of $1$ TeV and $2$ TeV, respectively.
The sharp drops around $m_X=2$ TeV and $4$ TeV respectively appear by the $X$ boson resonance in the annihilation processes.
End points of those curves are due to the constraints from DM direct search experiments.
In $\chi$ DM parameter space, only the vicinity of $s$-channel $X$ resonance is allowed, due to the stringent DM direct search bound.
Red solid (dashed) curve represent the predicted $N_\mathrm{eff}$ of $3.5 \, (3.1)$ by taking the contribution of $\nu_R$ radiation into account.
The LHC provides a more stringent limit for a lighter $m_X$ mass region $m_X \lesssim 3.5$ TeV, while the cosmological bound is significant for $m_X \gtrsim 3.5$ TeV.

For comparison, the different $x_H$ models are also displayed in Fig.~\ref{Fig:Constr}.
The upper-left panel shows the results for $x_H =1$.
In this case, both the LHC and cosmological $N_\mathrm{eff}$ bounds are more severe than those of the $U(1)_{B-L}$ case. Hence, a smaller $g_X$ than that in  the $U(1)_{B-L}$ model are allowed.
The LHC bound becomes most less stringent for $x_H=-1.2$~\cite{Okada:2016tci}, which is shown in the lower-left panel.
In this case, the $N_\mathrm{eff}$ constraint is comparable or more stringent than the LHC bound for all mass range of $m_X$ in the plot.  

Finally, we consider the case of $x_H =-2$, which corresponds to $U(1)_R$. 
This is a case that the constraint from direct DM search becomes weaker by the destructive interference between proton and neutron in the cross section, which can be easily seen by rewriting the SI WIMP-nucleon cross section (\ref{Eq:directCrossSection}) as 
\begin{equation}
\sigma_{SI} \propto (A (x_H+4)+2 x_H Z)^2.
\end{equation}
For atomic nucleus with $A \simeq 2 Z$, the cancellation occurs around $x_H \simeq -2$.
Thus, a larger $g_X$ than that in the $U(1)_{B-L}$ model is allowed.
As can be seen in the lower-right panel of Fig.~\ref{Fig:Constr},
 $m_X$ is not necessarily so close to $2 m_{\chi}$,
 and more interestingly, there exist parameter points that
 thermal DM with the mass of $2$ TeV and $\Delta N_\mathrm{eff} = \mathcal{O}(0.1)$ can be simultaneously realized.
We note that another model for the simultaneous realization of thermal DM and DR was proposed
 in Ref.~\cite{Okada:2019sbb},
 where DM is flavored $U(1)$ interacting scalar and the gauge boson is light.

\section{Summary}

We have proposed a simple extension of the SM with an extra $U(1)_X$ gauge symmetry 
with an alternative charge assignment for right-handed neutrinos.
The type-II seesaw mechanism is an economical way
 to generate appropriate neutrino masses in this framework
 and an additional Higgs doublet to have a trilinear interaction with the triplet Higgs field.
Since the tadpole term is proportional to the squared VEV of the additional doublet Higgs field, 
 the smallness of neutrino mass could be understood as the consequence of 
 the smallness of the additional Higgs VEV.
Thus, without introducing a very small dimensionful parameter in the Higgs potential, 
 we can observe the dilepton decay of the doubly-charged Higgs boson, which is 
 evidence of nonconservation of lepton number.

We have also derived the constraints for thermal DM and DR.
Since we have introduced only one SM singlet $U(1)_X$ breaking scalar,
it predicts the existence of one massless state and Dirac fermion in the right-handed neutrino sector. 
The LHC sets a stringent bound on the model parameters for $m_X \lesssim $ a few TeV, 
 while the DR constraint is more significant for $m_X \gtrsim $ a few TeV.
The most stringent bound comes from the DM physics in the wide range of parameter space.
The null results of direct DM searches impose the upper bound on the $U(1)_X$ gauge coupling.
A case with $x_H\simeq -2$ is exceptional because the direct DM search bound becomes weaker due to cancellation in nucleon DM scattering cross section.
Then, thermal DM and $\Delta N_\mathrm{eff} = \mathcal{O}(0.1)$ can be consistent.
Such an $N_\mathrm{eff}$ would be interesting from the viewpoint of Hubble tension.

Another important but unaddressed subject is baryogenesis, which is beyond the scope of this paper.
We, here, note that thermal leptogenesis by heavy-triplet Higgs
 particles~\cite{Ma:1998dx,Hambye:2003ka,Hambye:2005tk,Chongdar:2021tgm} or Affelck-Dine baryogenesis
 by scalar condensations~\cite{Barrie:2021mwi} appears to be promising.

\section*{Acknowledgments}
This work is supported in part by the U.S. DOE Grant No.~DE-SC0012447 (N.O.),
 the Japan Society for the Promotion of Science (JSPS) KAKENHI Grants
 No.~19K03860, No.~19K03865 and No.~21H00060 (O.S.).

\appendix

\section{Amplitude}
\label{sec:AppA}

%
We give explicit formulas of the invariant amplitude squared of scatterings between the SM fermions and $\nu_R$.

\begin{align}
&\int \overline{|\mathcal{M}(u\bar{u} \leftrightarrow \nu_R\nu_R)|^2}d\cos\theta = \frac{8 g^4 s \left(m_f^2 ((7 x_H+40) x_H+16)+s ((17 x_H+20) x_H+8)\right)}{27 \left(\Gamma_X^2 m_X^2+\left(m_X^2-s\right)^2\right)}, \\
&\int \overline{|\mathcal{M}(d\bar{d} \leftrightarrow \nu_R\nu_R)|^2}d\cos\theta = 
 \frac{8 g^4 s \left(m_f^2 (16-x_H (17 x_H+8))+s ((5 x_H-4) x_H+8)\right)}{27 \left(\Gamma_X^2 m_X^2+\left(m_X^2-s\right)^2\right)}, \\
&\int \overline{|\mathcal{M}(\ell\bar{\ell} \leftrightarrow \nu_R\nu_R)|^2}d\cos\theta =\frac{8 g^4 s \left(m_f^2 ((7 x_H+24) x_H+16)+s ((5 x_H+12) x_H+8)\right)}{3 \left(\Gamma_X^2 m_X^2+\left(m_X^2-s\right)^2\right)}, \\
&\int \overline{|\mathcal{M}(\nu\bar{\nu} \leftrightarrow \nu_R\nu_R)|^2}d\cos\theta =
\frac{8 g^4 s^2 (x_H+2)^2}{3 \left(\Gamma_X^2 m_X^2+\left(m_X^2-s\right)^2\right)} .
\end{align}
%



\end{document}